\documentclass{ws-ijmpe} 

\begin{document}

\markboth{Avraham Gal}{$\bar K$-Nuclear Deeply Bound States?}

\catchline{}{}{}{}{}
%%%%%%%%%%%%%%%%%%%%%%%%%%%%%%%%%%%%%%%%%%%%%%%%%%%%%%%%%%%%%%%%%%%% 

\title{$\bar K$-NUCLEAR DEEPLY BOUND STATES?
\footnote{Dedicated to Walter Greiner on the occasion of his 70th birthday, 
presented at the International Symposium on Heavy Ion Physics, ISHIP 2006, 
Frankfurt, April 2006.}} 

\author{\footnotesize AVRAHAM GAL}

\address{Racah Institute of Physics, The Hebrew University \\ 
Jerusalem 91904, Israel \\ avragal@vms.huji.ac.il} 

\maketitle 

\begin{history}
\received{(15 May 2006)}
%\revised{(revised date)}
\accepted{(xx yy 2006)}
%\comby{(xxxxxxxxxx)}
\end{history}

\begin{abstract}
Following the prediction by Akaishi and Yamazaki of relatively narrow 
$\bar K$-nuclear states, deeply bound by over 100 MeV where the 
main decay channel $\bar K N \to \pi \Sigma$ is closed, several experimental 
signals in stopped $K^-$ reactions on light nuclei have been interpreted 
recently as due to such states. In this talk I review (i) the evidence from 
$K^-$-atom data for a {\it deep} $\bar K$-nucleus potential, as attractive 
as $V_{\bar K}(\rho_0) \sim -(150 - 200)$ MeV at nuclear matter density, 
that could support such states; and (ii) the theoretical arguments for 
a {\it shallow} potential, $V_{\bar K}(\rho_0) \sim -(40 - 60)$ MeV. 
I then review a recent work by Mare\v{s}, Friedman and Gal in which 
$\bar K$-nuclear bound states are generated dynamically across the 
periodic table, using a RMF Lagrangian that couples 
the $\bar K$ to the scalar and vector meson fields mediating the nuclear 
interactions. The reduced phase space available for $\bar K$ absorption 
from these bound states is taken into account by adding a density- and 
energy-dependent imaginary term, underlying the corresponding $\bar K$-nuclear 
level widths, with a strength constrained by $K^-$-atom fits. Substantial 
polarization of the core nucleus is found for light nuclei, with central 
nuclear densities enhanced by almost a factor of two. The binding 
energies and widths calculated in this dynamical model differ appreciably 
from those calculated for a static nucleus. These calculations provide 
a lower limit of $\Gamma_{\bar K} \sim 50 \pm 10$ MeV on the width of nuclear 
bound states for $\bar K$ binding energy in the range 
$B_{\bar K} = 100 - 200$ MeV. 
\end{abstract}

\section{Introduction} 

The $\bar K$-nucleus interaction near threshold is strongly attractive and 
absorptive as suggested by fits to the strong-interaction shifts and widths 
in $K^-$-atom levels\cite{BFG97}. Global fits yield `deep' optical potentials 
$V_{\bar K}(\rho_0) \sim -(150-200)$ MeV \cite{FGB93,FGB94,FGM99,MFG06}, 
whereas other, more theoretically inclined works that fit the low-energy 
$K^-p$ reaction data, including the $I=0$ unstable bound state $\Lambda(1405)$ 
as input for constructing a density dependent optical potential, suggest 
relatively `shallow' potentials with $V_{\bar K}(\rho_0) \sim -(40-60)$ MeV 
\cite{SKE00,ROs00,BGN00,CFG01}. The issue of the depth of the attractive $\bar K$-nucleus potential is 
briefly reviewd in Sections \ref{sec:shallow},\ref{sec:deep}. 

\begin{figure}[th]
\centerline{\psfig{file=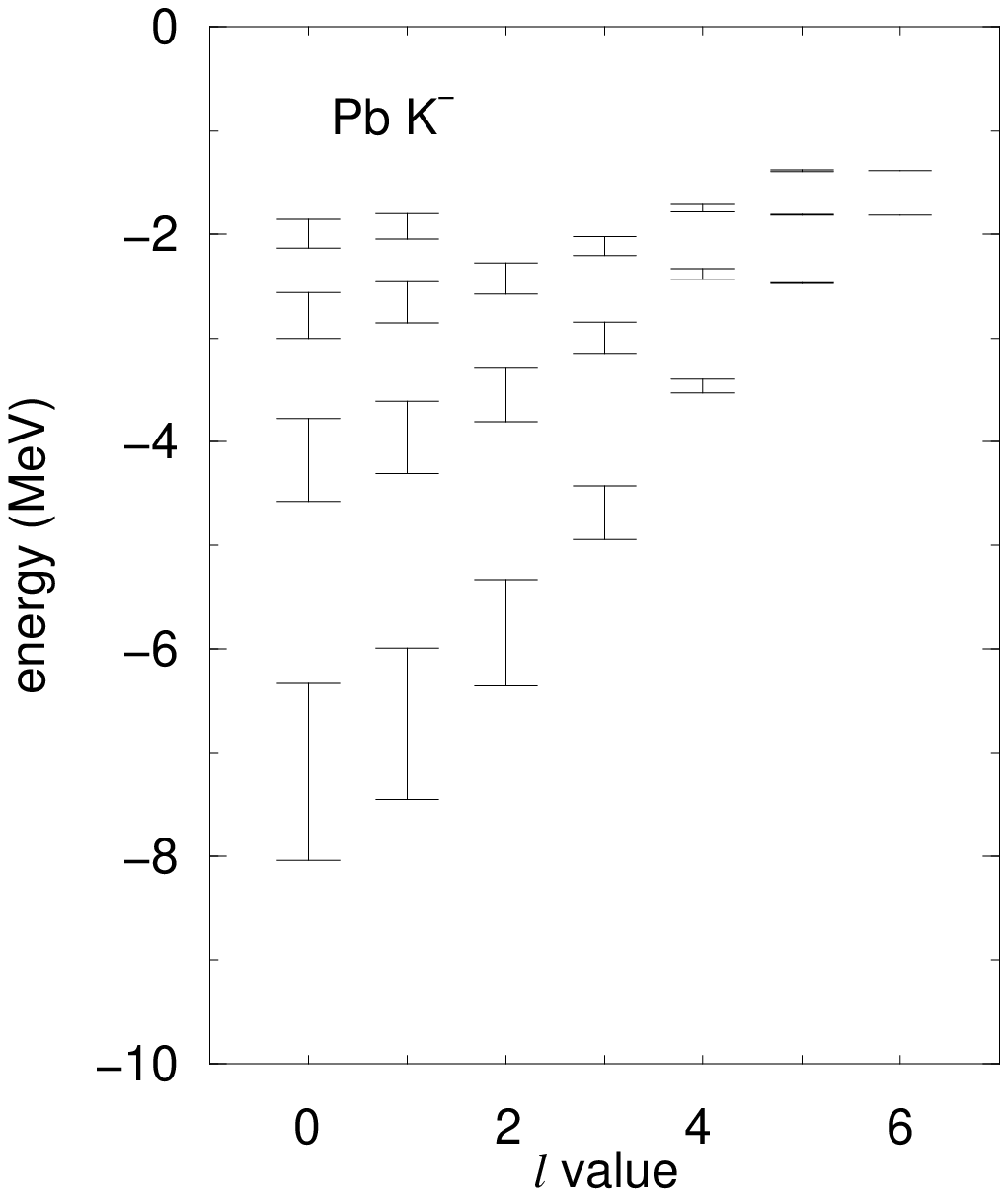,width=6.2cm}
\hspace*{3mm} 
\psfig{file=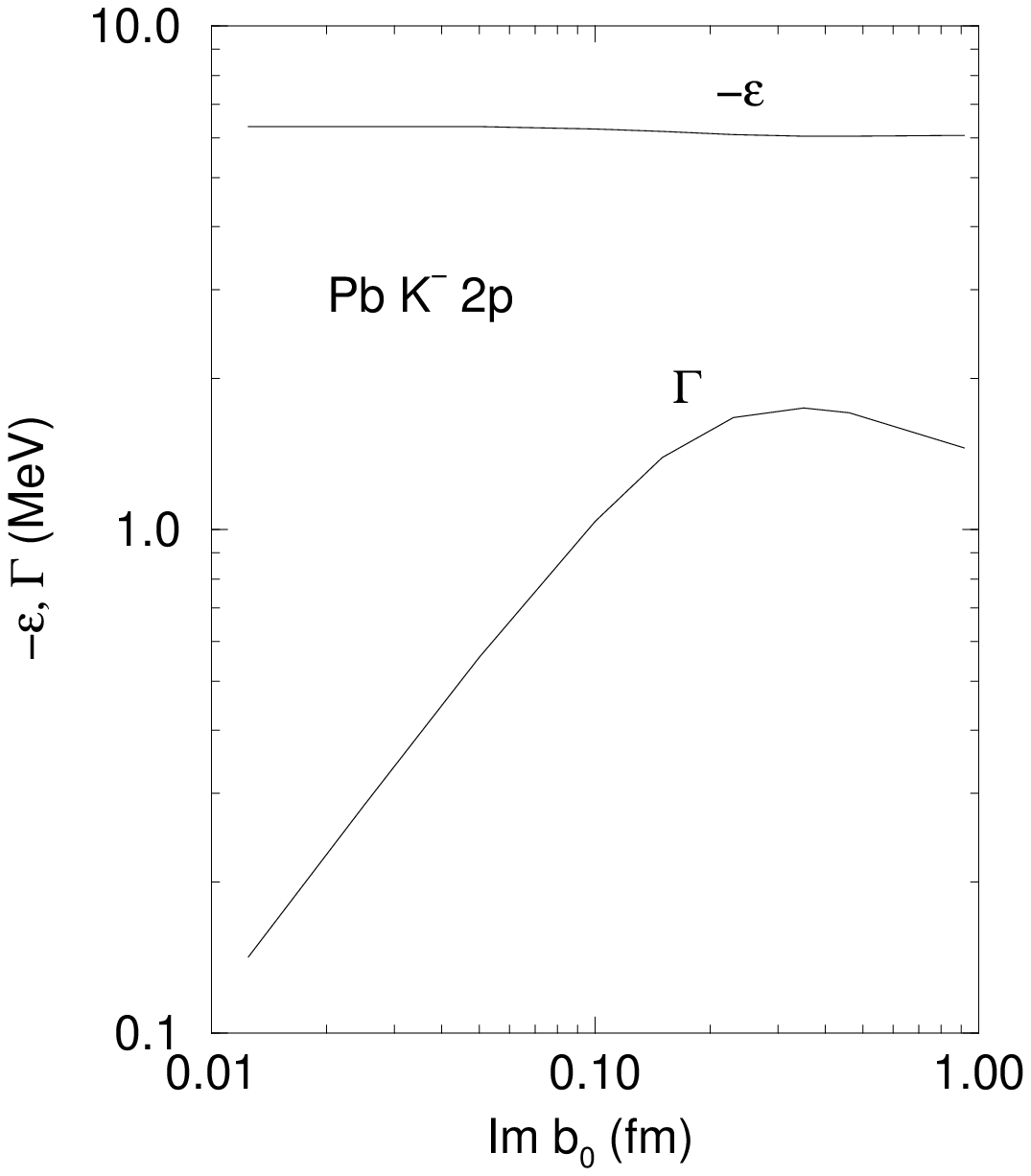,width=6.2cm}
}
\vspace*{8pt} 
\caption{Left: calculated spectrum of $K^-$ `deeply bound' atomic states 
in $^{208}$Pb. Right: saturation of width $\Gamma$ for the $2p$ 
$K^-$-$^{208}$Pb atomic state as function of Im~$b_0$, 
for Re~$b_0 = 0.62$~fm. \label{fig:KPb}}
\end{figure}

It is instructive at this introductory stage to mention that irrespective 
of the depth of the real part of the potential, and quite paradoxically due 
to its strong imaginary (absorptive) part, relatively narrow $K^-$ 
{\it deeply bound atomic states} were predicted to exist\cite{FGa99a,FGa99b}, 
also confirmed in Ref. \cite{BGN00}. 
Figure \ref{fig:KPb} from Ref. \cite{FGa99b} shows on the left-hand side 
a calculated spectrum of $K^-$ atomic states in $^{208}$Pb where, 
in particular, all the circular states below the $7i~(l=6)$ state are 
not populated by X-ray transitions due to the strong $K^-$-nuclear absorption, 
and on the right-hand side it demonstartes saturation of the $2p$ atomic-state 
width as a function of the absorptivity parameter Im~$b_0$ of the optical 
potential. The physics behind is that a strong imaginary part of $V_{\bar K}$ 
acts repulsively, suppressing the {\it atomic} wavefunction in the region of 
overlap with Im~$V_{\bar K}$. The calculated width of the `deeply bound' 
atomic $1s$ and $2p$ is less than 2 MeV, calling for experimental ingenuity 
how to form these levels selectively by a non radiative process\cite{FGa99c}. 
Relatively narrow $\bar p$ deeply bound atomic states were also predicted, 
due to the same width saturation mechanism\cite{FGa99b}. 

This saturation mechanism does not hold for 
nuclear states which retain very good overlap with the potential. Hence, 
the question to ask is whether it is possible at all to bind {\it strongly} 
$\bar K$ mesons in nuclei and are such bound states sufficiently narrow to 
allow observation and identification? 
The current experimental and theoretical interest in this question was 
triggered back in 1999 by the suggestion of Kishimoto\cite{Kis99} to look 
for such states 
in the nuclear reaction $(K^{-},p)$ in flight, and by Akaishi and 
Yamazaki\cite{AYa02} who suggested to look for a $\bar K NNN$ $I=0$ state 
bound by over 100 MeV for which the main $\bar K N \to \pi \Sigma$ decay 
channel would be kinematically closed. Some experimental evidence has been 
presented recently for candidate states in $(K^{-}_{\rm stop},n)$ and 
$(K^{-}_{\rm stop},p)$ reactions on $^4$He (KEK-PS E471,\cite{ISB03,SBF04} 
respectively), in the $(K^{-},n)$ in-flight reaction on $^{16}$O 
(BNL-AGS, parasite E930 \cite{KHA03}), and in $K^{-}_{\rm stop}$ reactions 
in Li and $^{12}$C, observing back-to-back $\Lambda p$ pairs from 
$K^- pp \to \Lambda p$ (DA$\Phi$NE-FINUDA \cite{ABB05}). None of it is 
sufficiently conclusive. 

It is interesting then to study theoretically $\bar K$ nuclear states 
in the range of binding energy $B_{\bar K} \sim 100 - 200$ MeV, 
as suggested by these recent experiments, and in particular the 
width anticipated for such deeply bound states. The relatively shallow 
$\bar K$-nucleus potentials\cite{BGN00,CFG01} derived 
from the microscopic construction by Ramos and Oset\cite{ROs00}, which are 
discussed in Section \ref{sec:shallow}, are of no use in this context, 
since they cannot generate even within a dynamical calculation binding 
energies substantially greater than the potential depth of about 50 MeV. 
One must therefore depart from the microscopic approach in favor 
of a more phenomenologically inclined model which is constrained 
by data other than two-body $\bar K N$ observables. The theoretical 
framework described here in Section \ref{sec:RMF}, due to Mare\v{s}, 
Friedman and Gal\cite{MFG06,MFG05}, is the relativistic mean field (RMF) 
model for a system of nucleons and one $\bar K$ meson interacting through 
the exchange of scalar ($\sigma$) and vector ($\vec {\omega},\vec {\rho}$) 
boson fields which are treated in the mean-field approximation. 
The RMF is a systematic approach used across the periodic table beyond 
the very light elements explored by other techniques, and it can be used 
also to study multi-$\bar K$ configurations and to explore the $\bar K$ 
condensation limit\cite{SGM94,SMi96}. Similar RMF calculations have been 
recently reported by the Frankfurt group for $\bar N$ states in 
nuclei\cite{BMS02,MSB04}.

\section{$\bar K$-nucleus potential from chirally motivated 
models} 
\label{sec:shallow} 

\begin{figure}[th] 
\centerline{\psfig{file=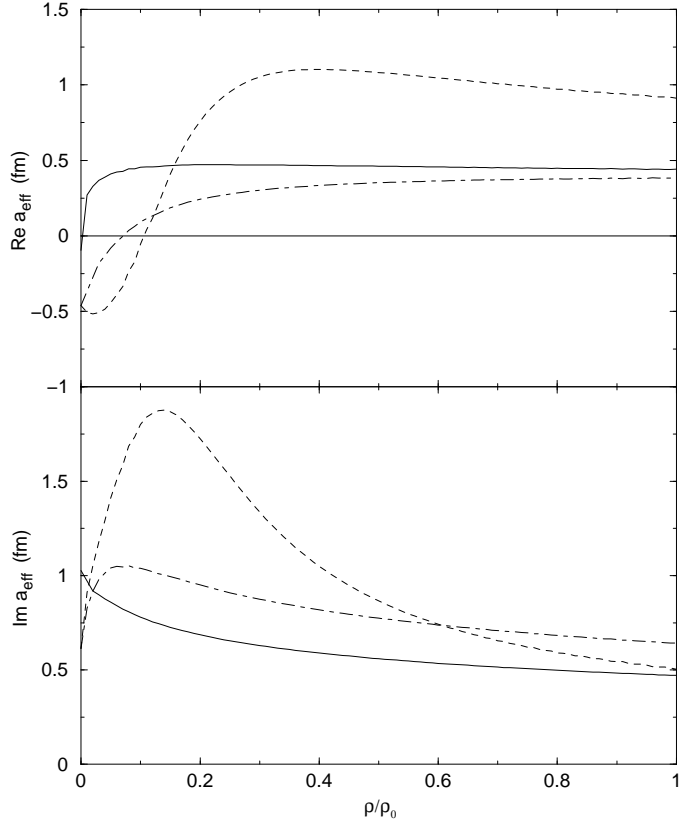,width=9cm}}
\vspace*{8pt}
\caption{Re~$a_{\rm eff}$ (top) and Im~$a_{\rm eff}$ (bottom) for the 
$\bar K N$ scattering length as function of the density, calculated without 
requiring self consistency (dashed) and requiring it (dot-dash and solid).} 
\label{fig:SC} 
\end{figure}

The Born approximation for the $\bar K$-nucleus potential due to the 
leading-order Tomozawa-Weinberg (TW) vector term of the chiral effective 
Lagrangian \cite{WRW97}, 
\begin{equation} 
\label{eq:chiral} 
V_{\bar K}~ =~ - ~\frac{3}{8f_{\pi}^2}~\rho 
\end{equation} 
where $f_{\pi} \sim 93$ MeV is the pseudoscalar meson decay constant, 
yields sizable attraction $V_{\bar K}(\rho_0) \sim -55$ MeV 
for $\rho _0 = 0.16$ fm$^{-3}$. 
Iterating the TW term plus next-to-leading-order terms, 
within an {\it in-medium} coupled-channel approach constrained 
by the $\bar K N - \pi \Sigma - \pi \Lambda$ data near the 
$\bar K N$ threshold, roughly doubles this $\bar K$-nucleus attraction. 
It is found in these calculations 
(e.g. Ref. \cite{WKW96}) that the $\Lambda(1405)$ quickly dissolves in the 
nuclear medium at finite densities, well below $\rho _0$, so that 
the repulsive free-space scattering length $a_{K^-p}$ becomes 
{\it attractive}, and together with the weakly density dependent 
attractive $a_{K^-n}$ it yields an attractive density dependent 
effective isoscalar scattering length $a_{\rm eff}(\rho)$, denoted 
here also as $b_0(\rho)$ (positive for attraction): 
\begin{equation} 
\label{eq:b0}
b_0(\rho)={\frac{1}{2}}(a_{K^-p}(\rho)+a_{K^-n}(\rho))~,
~~~~b_0(\rho_0) \sim 0.9~{\rm fm}~, 
\end{equation} 
leading to a strongly attractive $\bar K$-nucleus optical potential: 
\begin{equation} 
\label{eq:trho} 
V_{\bar K}(r)~ = ~-{\frac{2\pi}{\mu_{KN}}}b_0(\rho)\rho(r)~, 
~~~~{\rm Re}~V_{\bar K}(\rho_0) \sim -110~{\rm MeV}. 
\end{equation} 
Here $\mu_{KN}$ is the $\bar K N$ reduced mass. 
However, when $V_{\bar K}$ is calculated {\it self consistently} 
(SC), namely by including $V_{\bar K}$ in the in-medium propagator 
used in the Lippmann-Schwinger equation determining $V_{\bar K}$ 
as a $\bar K$-nucleus T matrix, the resultant $\bar K$-nucleus potential is 
only moderatelly attractive, with depth between $40 - 60$~MeV. 
This is shown in Fig. \ref{fig:SC}, where the reduction of Re~$a_{\rm eff}$ 
from the (non SC) dashed line to the (SC) solid line is clearly 
seen\cite{CFG01}. The reason for this weakening of the T matrix, 
approximately back to its TW potential kernel Eq. (\ref{eq:chiral}), 
is the strong absorptive effect of $V_{\bar K}$ in the $\bar K$-nucleus 
propagator which suppresses the higher-order TW potential. 
It is the main reason that different theoretical calculations 
that obtain $\bar K$-nucleus potential depths $\sim 100$~MeV 
without requiring SC, find ultimately upon requiring SC depths 
$\sim 50 \pm 10$~MeV at $\rho_0$\cite{SKE00,ROs00,CFG01}. 
The Akaishi and Yamazaki scheme of calculation\cite{AYa02} differs in many 
respects from the theoretical derivations here outlined. 
It gets rid of the $\pi \Sigma$ coupled-channel effects too early 
in the calculation, essentially using a one-channel ${\bar K} N$ 
energy-independent complex potential that fits the $I=0,1$ scattering lengths 
and the position of the $\Lambda(1405)$ unstable bound state for $I=0$. 
This kind of construction is far from being unique, certainly with respect 
to models that use a comprehensive set of ${\bar K} N$ low-energy 
data\cite{ROs00,CFG01}. Furthermore, their analog of the TW potential kernel 
is considerably more attractive; and moreover, they do not impose the SC 
requirement, and it appears that the absorptive effect mentioned above 
is missing in their scheme. Therefore, their calculations have little 
common with the other `theoretical' calculations in this field. 

\begin{figure}[th] 
\centerline{\psfig{file=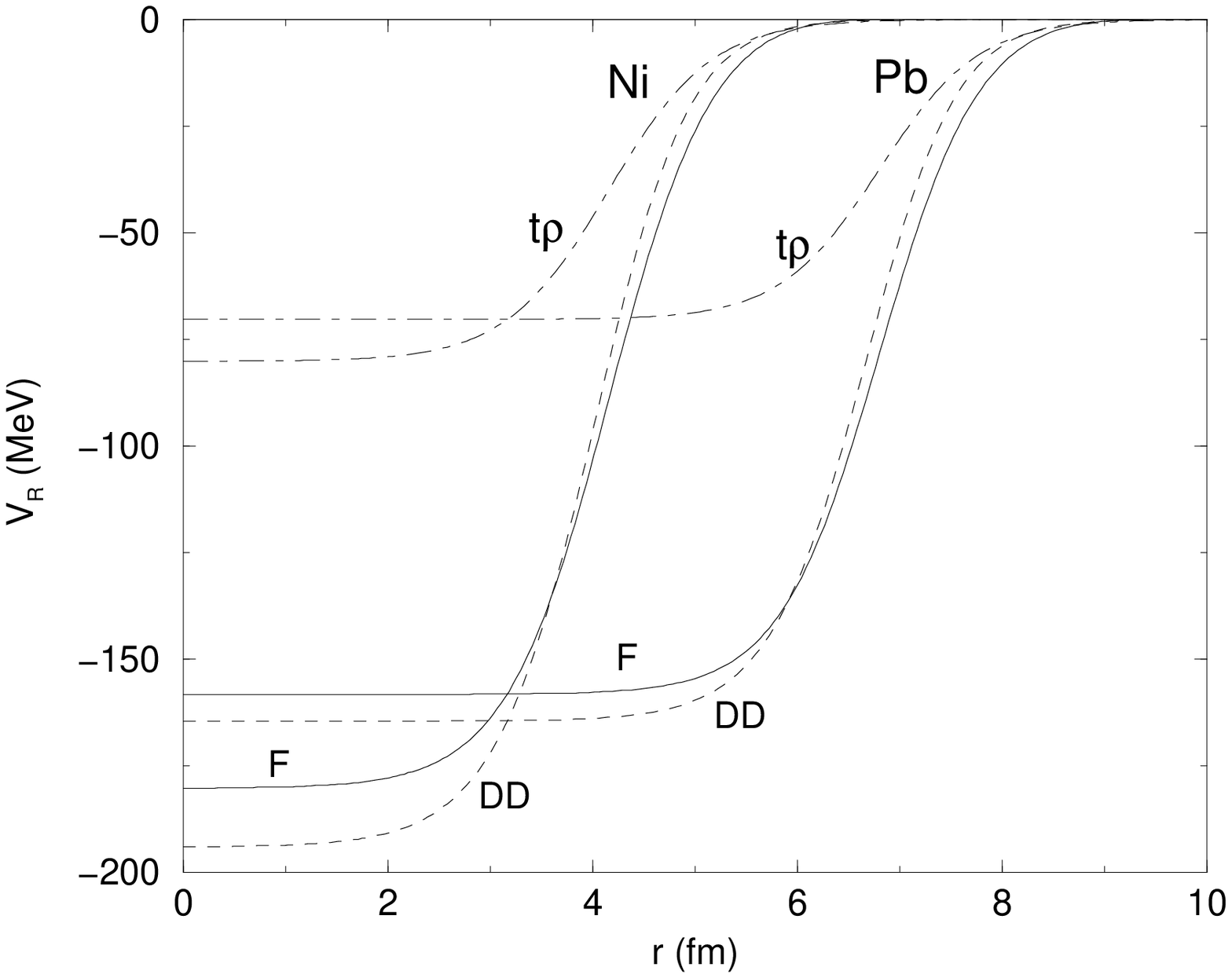,width=6.2cm}
\hspace*{3mm}
\psfig{file=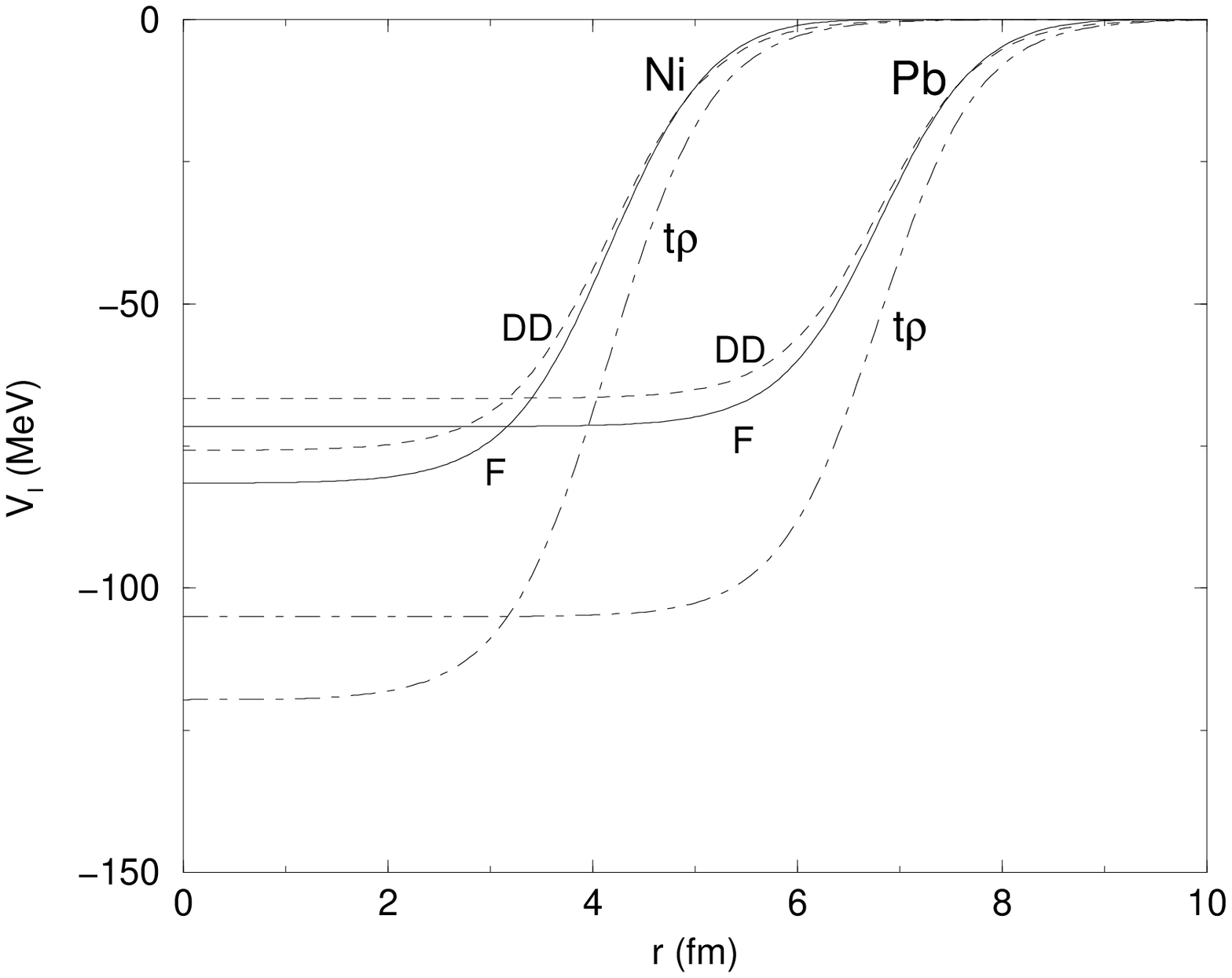,width=6.2cm}
}
\vspace*{8pt}
\caption{Real part (left) and imaginary part (right) of the 
$\bar K$-nucleus potential for $^{58}$Ni and $^{208}$Pb, 
obtained in a global fit to $K^-$-atom data, for a $t\rho$ 
potential, for the DD potential{\protect \cite{BFG97}} and for potential 
F {\protect \cite{MFG06}}, see text.} 
\label{fig:V} 
\end{figure}

\section{$\bar K$-nucleus potential from fits to $K^-$-atom data} 
\label{sec:deep} 

The $K^-$-atom data used in global fits\cite{BFG97} span a range
of nuclei from $^7$Li to $^{238}$U, with 65 level-shift and -width 
data points. Figure \ref{fig:V} shows the real part and the imaginary part 
of the $\bar K$-nucleus potential for $^{58}$Ni and $^{208}$Pb for 
a $t\rho$ potential, where the complex strength $t$ is fitted to these data, 
and for two density-dependent potentials marked by DD and F, also fitted to 
the same data. For the real part, the depth of the 
$t\rho$ potential for a typical medium-weight to heavy nucleus is about 
80~MeV, whereas the density-dependent potentials are considerably deeper, 
$150-200$~MeV. These latter potentials also yield substantially lower 
$\chi ^2$ values of 103 and 84, respectively, than the value 129 for the 
$t\rho$ potential. In particular, the shape of potential F departs 
appreciably from  $\rho (r)$ for $\rho (r)/\rho_0 \lesssim 0.2$, where the 
physics of the $\Lambda(1405)$ still plays a major role. We note that, 
although the two density-dependent potentials have very different 
parameterizations, the resulting potentials are quite similar. 
In model F, considered in Ref. \cite{MFG06}, one divides configuration space 
into an `internal' region and an `external' region for both Re~$V_{\bar K}$ 
and Im~$V_{\bar K}$: 
\begin{equation} 
\label{eq:DDF} 
b_0~\rightarrow ~B_0~F(r)~+~b_0~[1~-~F(r)]~~, 
\end{equation} 
using the function 
\begin{equation}
\label{eq:F}
F(r)~=~\frac{1}{e^x +1}~,~~~x=(r-R_x)/a_x~,~~~R_x=R_{x0}A^{1/3}+\delta _x~.
\end{equation}
Here the parameter $b_0$ represents the interaction in the external 
region, and may be held fixed at its free $\bar K N$ value, whereas 
the parameter $B_0$ represents the interaction inside the nucleus. 
The $K^-$ atomic fit F used a fixed value $a_x = 0.4$~fm, with little 
sensitivity for the precise value in the range $0.2- 0.5$~fm, optimizing 
for $R_{x0}=1.30 \pm 0.05$~fm, $\delta _x=0.8 \pm 0.3$~fm, thus 
implying that the modification of the free $\bar K N$ interaction, 
namely the transition from the $b_0$ term to the $B_0$ term in 
Eq. (\ref{eq:DDF}), takes place at radii somewhat outside of the nuclear 
`half-density' radius. Consequently, $V_{\bar K}$ departs in shape 
from $\rho (r)$ for densities less than about 20\% of the central density.

\section{$\bar K$-nucleus RMF Dynamical Calculations} 
\label{sec:RMF} 

\subsection{$\bar K$-nucleus RMF methodology} 

In this model, expanded in Ref. \cite{MFG06}, the (anti)kaon interaction with 
the nuclear medium is incorporated by adding to ${\cal L}_N$ the Lagrangian 
density ${\cal L}_K$ \cite{SGM94,SMi96}: 
\begin{equation}
\label{eq:Lk}
{\cal L}_{K} = {\cal D}_{\mu}^*{\bar K}{\cal D}^{\mu}K -
m^2_K {\bar K}K
- g_{\sigma K}m_K\sigma {\bar K}K\; .
\end{equation} 
The covariant derivative
${\cal D_\mu}=\partial_\mu + ig_{\omega K}{\omega}_{\mu}$ describes
the coupling of the (anti)kaon to the vector meson $\omega$.
The coupling of the (anti)kaon to the isovector $\rho$ meson was 
neglected, a good approximation for the light $N=Z$ nuclei.  
Whereas extending the nuclear Lagrangian ${\cal L}_N$ by the Lagrangian
${\cal L}_K$ does not affect the original form of the corresponding Dirac 
equation for nucleons, the presence of $\bar K$ leads to additional source 
terms in the equations of motion for the meson fields $\sigma$ and $\omega_0$ 
to which the $\bar K$ couples. The $\bar K$ thus affects the scalar and vector 
potentials which enter the Dirac equation for nucleons, and this leads to a 
rearrangement or polarization of the nuclear core.

In order to preserve the connection to studies
of $K^-$ atoms, the Klein Gordon (KG) equation of motion for the
$\bar K$ is written in the form\cite{CFG01}:
\begin{equation}
\label{eq:KG1}
\left[\Delta - 2{\mu}(B^{\rm s.p.}+V_{\bar K}+V_c) + 
(V_c+B^{\rm s.p.})^2 \right]{\bar K} = 0~~ ~~
(\hbar = c = 1), 
\end{equation}
where $V_c$ is the $K^-$ static Coulomb potential and 
$\mu$ is the $\bar K$-nucleus reduced mass. 
The superscript s.p. in
$B^{\rm s.p.}=B_{\bar K}^{\rm s.p.}+{\rm i}{\Gamma_{\bar K}}/2$
stands for the single-particle $\bar K$ binding energy which is equal
to the $\bar K$ separation energy only in the static calculation. 
The difference $B_{\bar K}^{\rm s.p.} - B_{\bar K}$ provides a measure of 
the nuclear rearrangement energy. 
The real part of the $\bar K$ optical potential $V_{\bar K}$ 
is then given by 
\begin{equation}
\label{eq:VOP1}
{\rm Re}\;V_{\bar K}={\frac {m_K}{\mu}} [{\frac{1}{2}}S - 
(1 - \frac{B_{\bar K}^{\rm s.p.} + V_c}{m_K})V - {\frac{V^2}{2m_K}}] \; ,
\end{equation} 
where $S = g_{\sigma K}\sigma$ and $V = g_{\omega K}\omega_0$ are 
scalar and vector potentials due to the $\sigma$ and $\omega$ mean fields,
respectively. Whereas Re~$V_{\bar K}$ is implicitly {\it state dependent} 
through the dynamical density dependence of the mean-field potentials $S$ 
and $V$, as expected from a RMF calculation, it is here also explicitly 
state dependent through 
the $[1 - (B_{\bar K}^{\rm s.p.} + V_c)/m_K]$ energy-dependent 
factor multiplying the vector potential $V$. 

Since the RMF approach does not address the imaginary part of the 
potential, ${\rm Im}~V_{\rm opt}$ was taken in a phenomenological $t\rho$ 
form, where its depth was fitted to the $K^-$ atomic data\cite{FGM99} 
with Im~$b_0$ = 0.62~fm. 
Note that the density $\rho$ in the present calculations is no longer
a static nuclear density, but is a {\it dynamical} entity affected by the
$\bar K$ interacting with the nucleons via boson fields. 
Suppression factors multiplying ${\rm Im}~V_{\rm opt}$ were introduced 
from phase-space considerations, taking into account the binding energy 
of the antikaon for the initial decaying state, and assuming two-body 
final-state kinematics for the decay products in the $\bar K N \to \pi Y$ 
mesonic modes as well as in the $\bar K N \to Y N$ nonmesonic modes.  

The coupled system of equations for nucleons and for the electromagnetic
vector field $A_0$, for the $\rho$ meson mean field, and for the mean fields 
$\sigma$ and $\omega_0$ with terms explicitly due to $\bar K$ coupling, 
as well as the KG equation (\ref{eq:KG1}) for $K^-$, were solved 
self-consistently using an iterative procedure. 
In order to produce different values of binding energies, 
$g_{\sigma K}$ was scaled down successively from its initial value 
$g^{\rm atom}_{\sigma K}$ appropriate to the $K^-$-atom fit and, 
once it reached zero, $g_{\omega K}$ too was scaled down until the $K^-$ 
$1s$ state became unbound. Obviously, the good {\it global} fit to the 
atomic data is thus lost. 
Furthermore, in order to scan the region of large values of $B_{K^-}$, 
of order 200~MeV, we also scaled up $g_{\sigma K}$ from its initial value, 
while keeping $g_{\omega K}$ fixed at $g^{\rm atom}_{\omega K}$. 
To give rough idea, whereas the static calculation gave $B_{K^-} = 132$~MeV 
for the $K^-$ $1s$ state in $^{12}$C, 
using $g^{\rm atom}_{\omega K},g^{\rm atom}_{\sigma K}$, 
the dynamical calculation gave $B_{K^-} = 172$~MeV for this same state. 

\subsection{Results for binding energies and widths}

\begin{figure}[th] 
\centerline{\psfig{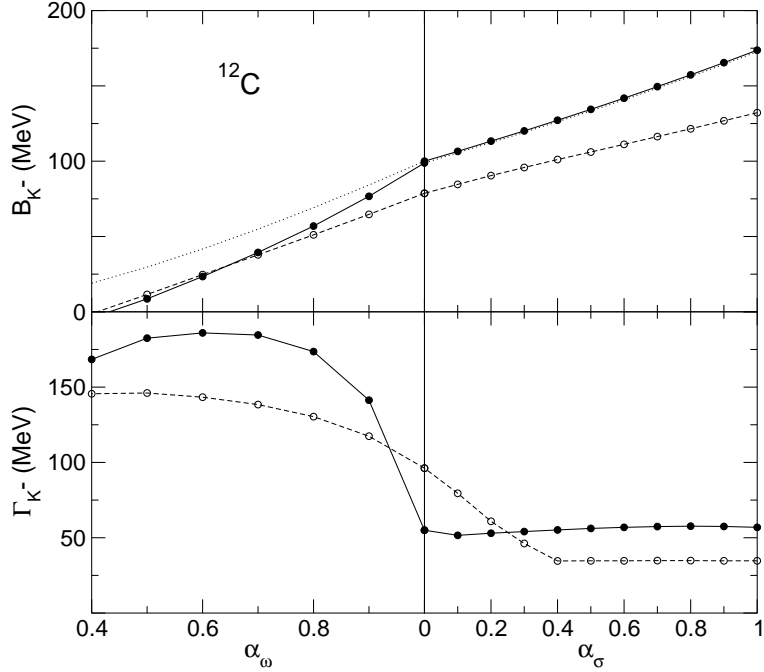}} 
\vspace*{8pt} 
\caption{$1s$ $K^-$ binding energy and width in $^{~~12}_{K^-}$C 
calculated statically (dashed lines) and dynamically (solid lines)
for the nonlinear RMF model NL-SH {\protect \cite{SNR93}} as function of
$\alpha_{\omega} = g_{\omega K}/g^{\rm atom}_{\omega K}$ and 
$\alpha_{\sigma} = g_{\sigma K}/g^{\rm atom}_{\sigma K}$: 
$\alpha_{\omega}$ is varied in the left panels as indicated, 
with $\alpha_{\sigma}=0$, and $\alpha_{\sigma}$ is varied in the right 
panels as indicated, with $\alpha_{\omega}=1$. The dotted line shows 
the calculated binding energy when the absorptive $K^-$ potential is 
switched off in the dynamical calculation.}
\label{fig:dynam}
\end{figure}

\begin{figure}[th]
\centerline{\psfig{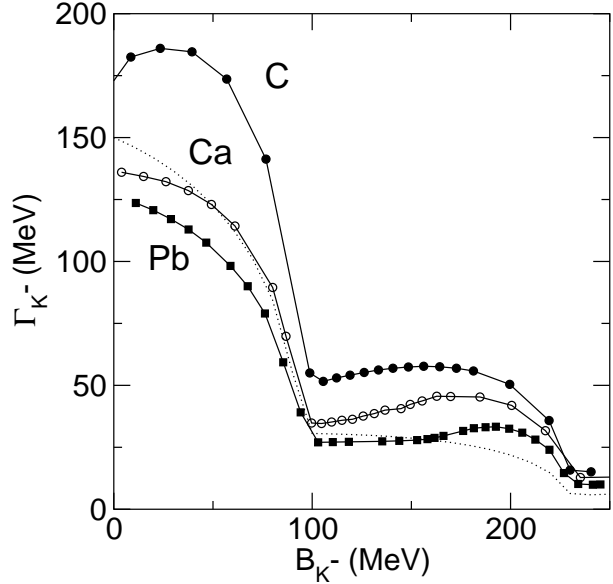}}
\vspace*{8pt} 
\caption{Dynamically calculated widths of the $1s$ $K^-$-nuclear state
in $^{~~12}_{K^-}$C, $^{~~40}_{K^-}$Ca and $^{~~208}_{K^-}$Pb 
as function of the $K^-$ binding energy for nonlinear RMF models. 
The dotted line is for a static nuclear-matter calculation with 
$\rho_0=0.16~{\rm fm}^{-3}$.}
\label{fig:Gamma}  
\end{figure}

A comparison between 
the statically calculated (dashed lines) and the dynamically calculated
(solid lines) $B_{K^-}$ and $\Gamma_{K^-}$ for the $1s$ state in
$^{~~12}_{K^-}$C is shown in Fig. \ref{fig:dynam} as a function of the
coupling-constant strenghts . We note that the width calculated 
dynamically for the $1s$ nuclear state in $^{~~12}_{K^-}$C does not fall 
below 50 MeV for the range of variation shown, whereas the corresponding 
limiting value of the statically calculated width is about 35 MeV. Another 
feature shown in Fig. \ref{fig:dynam} concerns the effect of the imaginary 
potential on the binding energy: the dynamically calculated binding energy 
$B_{K^-}$ when Im~$V_{\bar K}$ is switched off is shown by the dotted line. 
It is clear that the absorptive potential Im~$V_{\bar K}$ acts 
{\it repulsively} and its inclusion leads to less binding, particularly 
at low binding energies. The repulsive 
effect of Im~$V_{\bar K}$ gets weaker with $B_{K^-}$, along with the action 
of the kinematical phase-space suppression factors, and beginning with 
$B_{K^-} \sim 100$~MeV it hardly matters for the calculation of $B_{K^-}$ 
whether or not Im~$V_{\bar K}$ is included. 
The $A$ dependence of these features is considered in Ref.~\cite{MFG06}, 
leading to the following conclusions beginning approximately with $^{12}$C: 
\begin{itemize}
\item 
The $\bar K$ binding energy saturates, except for a small increase due to 
the Coulomb energy. 
\item 
The difference between the binding energies calculated dynamically and 
statically, $B_{\bar K}^{\rm dyn} - B_{\bar K}^{\rm stat}$, is substantial 
in light nuclei, increasing with $B_{\bar K}$ for a given value of $A$, and 
decreasing monotonically with $A$ for a given value of $B_{\bar K}$. 
It may be neglected only for very heavy nuclei. The same holds for the 
nuclear rearrangement energy $B_{\bar K}^{\rm s.p} - B_{\bar K}$ which is 
a fraction of $B_{\bar K}^{\rm dyn} - B_{\bar K}^{\rm stat}$. 
\item 
The width $\Gamma (B_{\bar K})$ decreases monotonically with $A$, as shown 
in Fig. \ref{fig:Gamma}. 
\end{itemize} 

Figure \ref{fig:Gamma} shows calculated
widths $\Gamma_{K^-}$ as function of the binding energy $B_{K^-}$ for 
$1s$ states in $^{~~12}_{K^-}$C and $^{~~40}_{K^-}$Ca, using the nonlinear 
NL-SH version\cite{SNR93} of the RMF model, and in 
$^{~~208}_{K^-}$Pb using the NL-TM1 version\cite{STo94}. 
The dotted line shows the static `nuclear-matter' limit
\begin{equation}
\label{eq:Gamma}
\Gamma_{K^-}~=~\frac{f}{1-\frac{B_{K^-}}{m_K}}~\Gamma_{K^-}^{(0)}~~, 
\end{equation}
where $f$ is the phase-space suppression factor used in the dynamical 
calculations, and $\Gamma_{K^-}^{(0)}$ is given by 
\begin{equation}
\label{eq:rho0}
\Gamma_{K^-}^{(0)}~=~{\frac{4\pi}{\mu_{KN}}}~{\rm Im}~b_0~\rho_0~~,
\end{equation}
for the static value ${\rm Im}~b_0=0.62$ fm used in the calculations and
for $\rho_0=0.16$ fm$^{-3}$. Eq. (\ref{eq:rho0}) holds for a $K^-$ 
Schroedinger wavefunction which is completely localized
within the nuclear central-density $\rho_0$ region. The additional factor
$(1-B_{K^-}/m_K)^{-1}$ in Eq. (\ref{eq:Gamma}) follows from using the KG
equation rather than the Schroedinger equation. 
It is clearly seen that the dependence of the width of the $K^-$ nuclear
state on its binding energy follows the shape of the dotted line
for the static nuclear-matter limit of $\Gamma_{K^-}$, Eq. (\ref{eq:Gamma}). 
This dependence is due primarily to the binding-energy
dependence of the suppression factor $f$ which falls off rapidly until
$B_{K^-} \sim 100$~MeV, where the dominant
$\bar K N \rightarrow \pi \Sigma$ gets switched off, and then stays
rather flat in the range $B_{K^-} \sim 100 - 200$~MeV where the width is 
dominated by the two-nucleon absorption modes. 
The larger values of width for the lighter nuclei are due to the 
dynamical nature of the RMF calculation, whereby the nuclear density is 
increased by the polarization effect of the $K^-$. The widths
calculated in the range $B_{K^-} \sim 100 - 200$~MeV assume considerably 
larger values than what the atatic calculation of the dotted line shows 
(except for Pb in the range $B_{K^-} \sim 100 - 150$~MeV). 
Adding perturbatively the residual width neglected in this calculation, 
mainly due to the $\bar K N \to \pi \Lambda$ secondary mesonic decay channel, 
a representative value for a lower limit $\Gamma_{\bar K} \sim 50 \pm 10$~MeV 
holds in the binding energy range $B_{K^-} \sim 100 - 200$~MeV.

\subsection{Nuclear polarization effects} 

\begin{figure}[th]
\centerline{\psfig{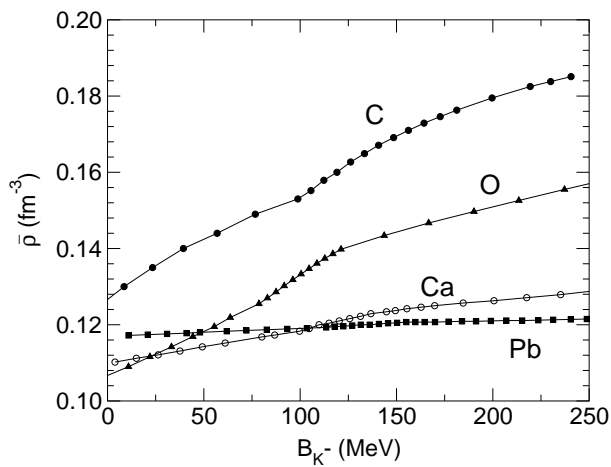}} 
\vspace*{8pt} 
\caption{Calculated average nuclear density $\bar \rho$ for 
$^{~~12}_{K^-}$C, $^{~~16}_{K^-}$O, $^{~~40}_{K^-}$Ca and
$^{~~208}_{K^-}$Pb as function of the $1s$ $K^-$ binding energy 
for the NL-SH RMF model {\protect \cite{SNR93}}. \label{fig:rhobar}} 
\end{figure} 

Figure \ref{fig:rhobar} shows the calculated average nuclear density 
$\bar \rho = \frac{1}{A}\int\rho^2d{\bf r}$ as a function of $B_{K^-}$ for 
the same $K^-$ nuclear $1s$ states as in Fig. \ref{fig:Gamma} and for $1s$ 
states in $^{~~16}_{K^-}$O. 
The average nuclear density $\bar \rho$ increases substantially in the 
light $K^-$ nuclei, for the binding-energy range shown here, to values 
about 50\% higher than for these nuclei in the absence of the $K^-$ meson. 
The increase of the central nuclear densities is even bigger, 
as demonstrated in the next figure, but is confined to a small region 
of order 1 fm from the origin. 

\begin{figure}[th]
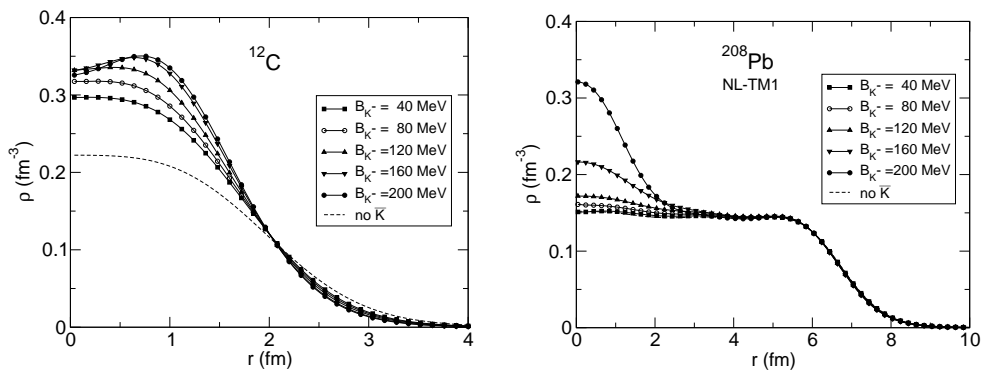
 
\centerline{\psfig{file=K05fig7.eps,width=6.2cm}
\hspace*{3mm}
\psfig{file=K06PbTM1rho.eps,width=6.2cm}
}
\vspace*{8pt}
\caption{Calculated nuclear density $\rho$ of $^{~~12}_{K^-}$C (left) 
and of $^{~208}_{K^-}$Pb (right) for several $B_{K^-}$ values of $1s$ $K^-$ 
nuclear states, using the nonlinear RMF models NL-SH {\protect \cite{SNR93}} 
and NL-TM1 {\protect \cite{STo94}}, respectively.} 
\label{fig:rho} 
\end{figure} 

Figure \ref{fig:rho} shows on the left-hand side calculated nuclear 
densities of $^{~~12}_{K^-}$C for several values of the $1s$ $K^-$ binding 
energies. The purely nuclear density, in absence of the $K^-$ meson, 
is given by the dashed curve. 
The maximal value of the nuclear density is increased by 
up to $75\%$ in the range of binding energies spanned in the figure, 
and the enhancement is close to uniform over the central region of 
$r \lesssim 1$~fm, decreasing 
gradually to zero by $r=2$~fm which already marks the nuclear surface. 
In this fairly small nucleus, the density is enhanced over a substantial 
portion of the nucleus. This is different than in heavier nuclei, 
as shown on the right-hand side of Fig. \ref{fig:rho} for $^{~208}_{K^-}$Pb. 
Here, the enhancement of the maximal density (at the center of the nucleus) 
is appreciable only for $B_{K^-}$ values above 150 MeV, 
reaching a factor of two, but it subsides almost completely by 
$r=2$~fm which is still well within the nuclear volume. 
As a result, the {\it average} nuclear density of $^{~208}_{K^-}$Pb 
shown in Fig. \ref{fig:rhobar} is only weakly enhanced as function 
of $B_{K^-}$. 

\begin{figure}[th] 
\centerline{\psfig{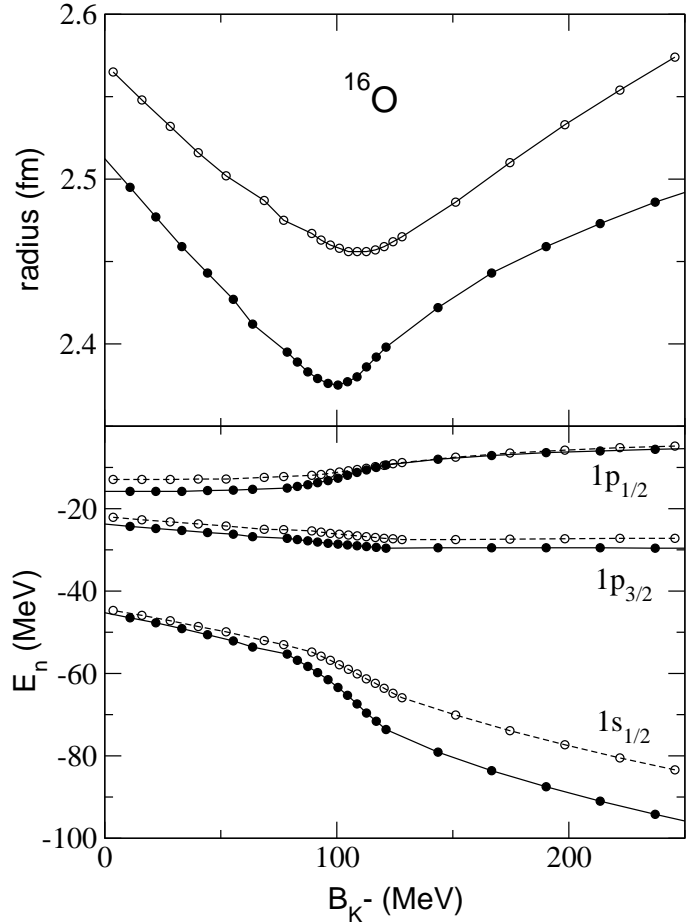}}
\vspace*{8pt}
\caption{Nuclear rms radius and neutron single-particle energies
for $^{~~16}_{K^-}$O as function of the $1s$ $K^-$ binding energy,
for the linear RMF model L-HS {\protect \cite{HSe81}} (open circles) and the
nonlinear RMF model NL-SH {\protect \cite{SNR93}} (solid circles).} 
\label{fig:nuclO}
\end{figure}

The upper and lower panels of Fig. \ref{fig:nuclO} show the calculated
nuclear rms radius and the $1s$ and $1p$ neutron single-particle energies
$E_n$, respectively, for $^{~~16}_{K^-}$O as a function of $B_{K^-}$. 
It is clear that the polarization effect of the $1s$ $K^-$ bound state on 
the $1s$ nuclear core is particularly strong. 
The differences between the linear and nonlinear models reflect the different 
nuclear compressibility and the somewhat different nuclear sizes obtained in 
the two models. The increase in the
nuclear rms radius of $^{~~16}_{K^-}$O for large values of $B_{K^-}$ is
the result of the reduced binding energy of the $1p_{1/2}$ state,
due to the increased spin-orbit term.

\section{Conclusions} 

In this talk I reviewed the phenomenological and theoretical evidence for 
a substantially attractive $\bar K$-nucleus interaction potential, from 
a `shallow' potential of depth $40-60$ MeV to a `deep' potential of depth 
$150-200$ MeV at $\rho_0$. I then reported on recent {\it dynamical} 
calculations\cite{MFG06} of deeply bound $K^-$ nuclear states across the 
periodic table, aimed at providing lower limits on the widths expected for 
binding energies in the range of $100-200$ MeV. Substantial polarization 
of the core nucleus was found in light nuclei. Almost universal dependence 
of $\bar K$ widths on the binding energy was found, for a given nucleus, 
reflecting the phase-space suppression factor on top of the increase provided 
by the density of the compressed nuclear cores. The present results already 
provide useful guidance for the interpretation of recent experimental results 
by placing a lower limit $\Gamma_{\bar K} \sim 50 \pm 10$ MeV. 
For lighter nuclear targets such as $^4$He, where the RMF approach becomes 
unreliable but where nuclear polarization effects are found larger using 
few-body calculational methods\cite{AYa02,DHA04}, one anticipates larger 
widths for $\bar K$ deeply bound states, if such states do exist.

\section*{Acknowledgements}

I wish to thank my collaborators Eli Friedman and Ji\v{r}\'{\i} Mare\v{s} 
for stimulating discussions, and Horst St\"{o}cker for supporting my 
participation in ISHIP 2006 through funds allocated by the 
Alexander von Humboldt Foundation. This work was supported in part by the 
Israel Science Foundation, Jerusalem, grant 757/05.

\end{document}